\documentclass{egpubl}
\usepackage{eurovis2025}

\WsSubmission      

\CGFccby

\usepackage[T1]{fontenc}
\usepackage{dfadobe}  

\usepackage{cite}  
\BibtexOrBiblatex
\electronicVersion
\PrintedOrElectronic
\ifpdf \usepackage[pdftex]{graphicx} \pdfcompresslevel=9
\else \usepackage[dvips]{graphicx} \fi

\usepackage{egweblnk}

\title[Beyond the Prototype]{Beyond the Prototype: Challenges of Long-Term Integration of Visual Analytics in Civic Spaces}

\author[M. jasim \& N. Mahyar]
{\parbox{\textwidth}{\centering M. Jasim$^{1}$\orcid{0000-0002-1250-3292}
        and N. Mahyar$^{2}$\orcid{0000-0003-1781-0029} 
        }
        \\
{\parbox{\textwidth}{\centering $^1$Louisiana State University, Baton Rouge, Louisiana, United States\\
         $^2$University of Massachusetts Amherst, Amherst, Massachusetts, United States
       }
}
}

\usepackage{balance}
\usepackage{xcolor}

\begin{document}


\maketitle
\begin{abstract}
Despite the recognized benefits of visual analytics systems in supporting data-driven decision-making, their deployment in real-world civic contexts often faces significant barriers. 
Beyond technical challenges such as resource constraints and development complexity, sociotechnical factors—including organizational hierarchies, misalignment between designers and stakeholders, and concerns around technology adoption hinder their sustained use. 
In this work, we reflect on our collective experiences of designing, developing, and deploying visual analytics systems in the civic domain and discuss challenges across design and adoption aspects. 
We emphasize the need for deeper integration strategies, equitable stakeholder engagement, and sustainable implementation frameworks to bridge the gap between research and practice.


\end{abstract}  

\section{Introduction}
Visual analytics systems are typically designed to facilitate the exploration of complex data, enabling data-driven analysis and supporting critical decision-making~\cite{munzner2014visualization}.
Despite their widely recognized benefits~\cite{cui2019visual}, the application and feasibility of visualization systems often remain limited to controlled studies within the research domain~\cite{burns2020evaluate}.
The transition of such systems from a controlled research environment to real-world deployment and sustained use faces major hurdles. 
While some challenges can be attributed to logistical issues, such as the resources and expertise needed to design, develop, host, and maintain software~\cite{andrienko2021big, wu2023grand}, in complex sociotechnical domains, many external challenges surface that inhibit the deployment, application, scalability, and sustained use of visual analytics systems. 



Digital civics is an interdisciplinary area exploring novel ways to utilize technology to promote democratic participation in the design and delivery of civic services~\cite{vlachokyriakos2016digital}. The goal is to enable wider participation by utilizing technology to support participatory democracy and greater transparency~\cite{corbett2018problem, kennethdd, Olivier}. 
While there has been a surge of interest from researchers and practitioners of HCI, digital civics, and visualization towards advancing technology to address issues of civic engagement and participatory democracy through visual analytics~\cite{faridani:2010, kriplean2012supporting, polis:2017, corbett2018problem, asad2017creating}, the civic domain is inherently complex due to the involvement of multiple stakeholders with conflicting perspectives~\cite{vlachokyriakos2016digital, corbett2018problem}, organizational constraints that impose difficult trade-offs~\cite{mahyar2019civic, jasimcommunityclick}, and large-scale public opinions often collected as unstructured data in diverse formats~\cite{mahyar2018communitycrit, jasim2021communityclick, jasimcommunityclick, asad2017creating}. 
In recent years, government agencies have increasingly turned to community-driven approaches to mitigate these issues. 
One approach has been to incorporate diverse stakeholder engagement and leverage technology for gathering and analyzing their input. 
Such engagements have shown the potential to foster more accurate and equitable decision-making while enhancing trust and legitimacy in governance~\cite{mahyar2018communitycrit, mahyar2016ud, jasim2021communitypulse}. 
However, even community-driven approaches are often top-down, where the public is predominantly involved as data providers and rarely included in the analysis and decision-making processes.

\begin{figure*}
    \centering
    \includegraphics[width=\linewidth]{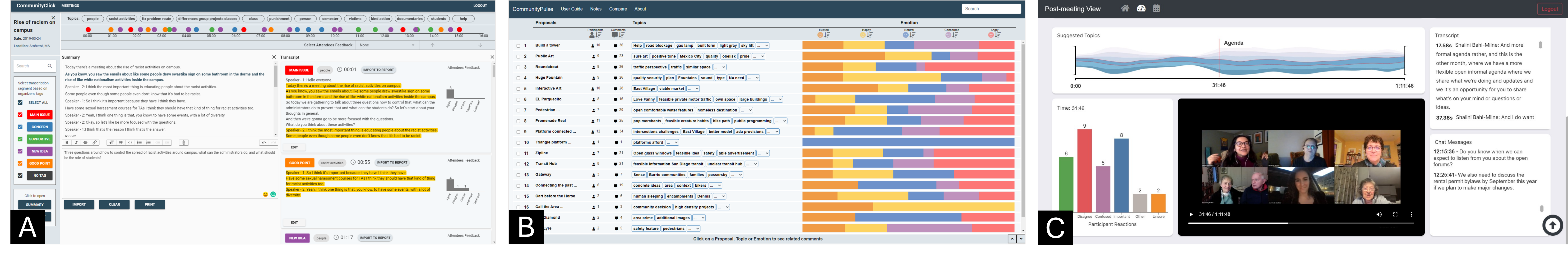}
    \caption{Three example visual analytics systems we designed, developed, and deployed in the civic domain with collaboration across diverse stakeholders: (A) CommunityClick~\cite{jasim2021communityclick}, (B) CommunityPulse~\cite{jasim2021communitypulse}, and (C) CommunityClick-Virtual~\cite{jasimcommunityclick}.}
    \label{fig:collage}
\end{figure*}

To mitigate such challenges, prior works in visualization and broader Human-Computer Interaction (HCI) have advocated for participatory and iterative user-driven design and development approaches, where end-users (often, the public) could be involved at all stages of creating visual analytics systems, from early design to implementation~\cite{hartson2012ux, kautz2010participatory}.  
Others have explored various developmental models for visualization applications that enable the flexibility and iterative nature of the visualization design process~\cite{munzner2009nested, lloyd2011human, isenberg2008grounded}. 
Such approaches have resulted in tools that enabled opinion sharing, consensus building, and data-driven decision-making in the civic domain~\cite{kriplean2012supporting, zilouchian2015procid, valkanova2014myposition}. 
For example, ConsiderIt enables the creation of pro-con lists to augment personal deliberation and help establish common grounds~\cite{kriplean2012supporting}. 
Procid enables consensus-building for distributed design decisions~\cite{zilouchian2015procid} by allowing diverse users to post new ideas and criteria, along with options to explore previous interactions. 
MyPosition allows users to vote on public policies and visualizes the results on public displays~\cite{valkanova2014myposition}. 
Despite the initial success, except for a few, many such systems did not achieve wide-scale adoption due to challenges that go beyond the design and development of visual analytics systems or the inclusion of stakeholders in participatory design. 
These challenges stem from divergent and often conflicting needs from a large variety of stakeholders~\cite{mahyar2019civic}, organizational complexities and hierarchies~\cite{corbett2018exploring}, and data privacy concerns~\cite{tomasello2023digital, mahyar2019civic}. 

We draw from over a decade of individual and collaborative experiences on designing, developing, and deploying building visual analytics systems with close collaboration with diverse stakeholders to discuss persistent challenges across both design and adoption aspects in the long-term adoption of visual analytics systems in the civic domain. 
We unravel challenges around building partnerships, mismatches among stakeholders' needs and expectations, and integrating explainable AI into visual analytics. 
We also emphasize the need to promote visual literacy across stakeholders, develop well-planned integration strategies, and implement equitable stakeholder engagement to address data privacy and ethical concerns to bridge the gap between research and practice.
By addressing these challenge, future work can contribute to more impactful and enduring applications of visual analytics in civic decision-making.

\section{Prior experiences}
In the last 10 years, we have designed, developed, and deployed several visual analytics systems in the civic domain. 
We draw on the experiences of working with diverse stakeholders, extensive design and development processes across many years of evolving programmers, resource allocation, and maneuvering administrative changes, and deployment of such systems in real-world scenarios. 

\subsection{Design and Development}
In collaboration with the town counselors and local government in the town of Amherst, Massachusetts, USA, we built and deployed multiple public engagement and visual analytics systems, including  CommunityClick~\cite{jasim2021communityclick} and CommunityClick-Virtual~\cite{jasimcommunityclick}. 
CommunityClick was designed to enhance physical town hall meetings by providing an augmented visualization-based feedback system, allowing attendees to express their opinions anonymously via iClickers. These inputs are integrated into a visual analytics dashboard, enabling organizers to generate data-driven summaries and reports that better reflect public sentiments.
CommunityClick-Virtual extends these capabilities to online public meetings, integrating multi-modal feedback mechanisms such as structured reactions and open-ended text responses. The system provides organizers with synchronized visual analytics that aggregate real-time audience feedback, transcripts, and discussion trends. 

For public input elicitation through microtask activities, Mahyar et al. developed CommunityCrit~\cite{mahyar2018communitycrit} in collaboration with civic leaders from San Diego, California, USA. 
The collected public input was then used in a visual analytics system, CommunityPulse~\cite{jasim2021communitypulse}, to surface public reaction, reflections on their own and others' ideas, and priorities by using visualization as a scaffolding for multi-level exploration and analysis of multifaceted community input. 
Furthermore, we designed and developed the UD Co-Spaces (Urban Design Collaborative Spaces)~\cite{mahyar2016ud}, which is a tabletop-centered multi-display environment to engage the public actively in collaborative urban design.  

\subsection{Deployment}
During the deployment of these systems, we closely collaborated with stakeholders, including the public, analysts, civic leaders, policymakers, and government officials. 
We deployed CommunityClick four times during public meetings in 2018 and 2019 in Amherst town halls and public meetings.
Shortly after these deployments, the COVID-19 pandemic severely disrupted civic processes across the United States, and all public engagement moved online. 
In response to this shift, we developed CommunityClick-Virtual, which was adopted by the Town of Amherst to facilitate hybrid town halls and public meetings in Amherst, Massachusetts, in 2021. 
CommunityClick-Virtual was also deployed several times to facilitate internal counselor meetings. 

In 2017, we partnered with a large planning group and community organizers in the city of San Diego, California, USA. 
With their close collaboration, we deployed CommunityCrit to gather people's input on a major redesigning of a busy cross-section in downtown San Diego for four weeks.
The collected data was used for CommunityPulse's deployment in 2019.

\section{Discussion}
The complex and intertwined nature of sociotechnical problems renders the design, development, deployment, and adoption of technologies built to address such issues inherently challenging.
We discuss these challenges across two dimensions: (1) Design challenges, where we shed light on challenges in designing and developing visual analytics systems involving diverse stakeholders, and (2) Adoption challenges, where we focus on the barriers of translation of research endeavors to sustained deployment and long-term adoption in the civic domain.  

\subsection{Design Challenges}
In discussing design challenges, we include barriers that manifest during the inception, design, prototyping, and eventual development of a feature-complete visual analytics system. 
These challenges extend beyond technological hurdles and encompass the complexities of wicked problems—highly interconnected and evolving issues with no clear-cut solutions. The wickedness arises from the intricate relationships between developers and stakeholders, the diverse traits and needs of the target audience, the influence of social and cultural norms, and underlying infrastructural constraints~\cite{rittel1973dilemmas, colding2019wicked}.

\subsubsection{Building Cross-Sector Partnerships}
To build effective visual analytics systems for the civic domain that can navigate the complex interplay among societal, political, economic, cultural, organizational, and computational factors, it is critical to undertake a human-centered approach~\cite{mahyar2019civic, baumer2022course}. 
To that end, it is imperative to build partnerships to include diverse stakeholder perspectives through co-design~\cite{steen2013co, niedderer2020working} to build features tailored to specific civic needs.
These stakeholders and end-users may include community organizations, local government, civic leaders, policymakers, and the public. 

While accessing and engaging such a wide variety of stakeholders demands significant efforts from developers, without stakeholders' representative involvement, the built system risks becoming biased towards certain groups and incapable of addressing the needs of other groups, leading to the inadvertent marginalization of ideas and requirements. 
Moreover, interests in partnership across these groups may vary~\cite{mahyar2019civic, jasim2021communitypulse}, making buy-in challenging and resulting in a lack of enthusiasm and engagement despite the best efforts from developers. 
Civic engagement is often voluntary and beyond the call of civic duty~\cite{jasim2021communityclick}; monetary or other incentives may not sway people's excitement to get involved.   
Overcoming these challenges calls for reimagining strategies to foster engagement and close collaboration among stakeholders, emphasizing shared goals and inclusive practices to inspire sustained engagement and meaningful impact~\cite{mahyar2019civic}.

\subsubsection{Mismatch Between Design Intent \& Stakeholder Needs}
The effective design and development of visualization-driven civic engagement and analytics systems require collaboration between computer scientists, analysts, policymakers, and community stakeholders. 
However, these stakeholders often have varying goals, priorities, expectations, and constraints, making consensus-building challenging. 
For instance, policymakers --- who make decisions based on public input --- often prefer summaries and high-level visual overviews to get a holistic understanding, and analysts might prefer drilling down to individual comments to gain a better understanding behind the rationale of public sentiments~\cite{mahyar2019civic, jasim2021communitypulse}.  
Furthermore, while government officials may be concerned about preserving data privacy, community organizations may prioritize data transparency and accessibility, and the public wants to see accountability and actions taken based on their input~\cite{jasim2021communityclick, jasimcommunityclick}.  

Such broad divergence has resulted in considering designing civic technologies as ``Wicked Problems''~\cite{rittel1973dilemmas, colding2019wicked} --- messy, unassuming, and often too complex with no trivial solutions.  
One approach to overcome these issues is to take a structured, participatory approach to bridge the gap between different stakeholders.
However, without strong communication and participatory frameworks, the roles of stakeholders, their involvement, and their impact on the design process remain undefined and underutilized~\cite{reynante2021framework}.
As such, it is imperative to develop early engagement and infrastructure to enable stakeholder roles and their involvement to ensure transparent, effective, equitable, and sustainable approaches in building visual analytics to address civic issues~\cite{reynante2021framework}. 

\subsubsection{Ensuring Equity and Inclusivity in Visual Analytics}

Many civic technologies assume universal access to the internet, smartphones, or computers, but many underserved communities lack reliable connectivity, devices, or digital literacy~\cite{jasim2021communityclick, liotta2023digitalization}. 
Challenges due to the assumption of access to technology often manifest in unexpected ways. 
For instance, hyperlinks to visual analytics systems can be disseminated through digital links or QR codes~\cite{jasimcommunityclick}. 
However, in situations where participants are provided with QR codes, they are expected to have smartphones with a digital camera and QR code reading software to use them. 
On the other hand, digital hyperlinks cannot be shared using posters or other manual approaches. 
Additionally, in situations where sharing the hyperlink is not possible, but QR codes can be used, it may force participants to work with the visual analytics system on smaller screens, despite having access to a larger screen on their desktop or laptop computers. 

These issues impact the design of visual analytics systems as the designers and developers have to put efforts to enable functionalities of the system across multiple devices and modalities. 
But perhaps a bigger challenge is in addressing language, cultural, and disability preferences to close the usability and digital literacy gaps of marginalized participants. 
Designing visual analytics systems in the civic domain that are equitable and accessible is essential to ensuring that all community members, regardless of digital literacy, disability, language, or socioeconomic status, can effectively engage with and benefit from these systems. 
This digital divide can be reduced by exploring approaches that enable low-technology accessible alternatives~\cite{cullen2001addressing}, companion applications that provide offline support~\cite{jasimcommunityclick}, and stateful implementation that can allow community members to save and restore their sessions for seamless engagement with visual analytics systems. 

\subsubsection{Embedding Explainable AI into Visual Analytics}


Historically, Artificial Intelligence (AI) algorithms have been considered and treated as a \textit{black box}~\cite{dwivedi2023explainable, kucher2022interdisciplinary}. 
Recent advancements in explainable AI (xAI) research have enabled a reduction in the opaqueness of these algorithms by making the AI models more transparent and understandable. 
However, they are often considerably more complex and difficult to visualize in a comprehensible way for non-experts, who constitute the majority of the audience in the civic domain~\cite{jasim2021communitypulse, burns2023we}.
This is especially pertinent in the digital civic domain when AI education, or lack thereof, can further increase the digital divide among the populace with access to such education~\cite{sieber2024civic}.
As such, integrating explainable AI components into visual analytics will require significant efforts to achieve the difficult task of balancing the need for simplicity and comprehensibility with the requirement for detailed and accurate explanations~\cite{baumer2022course}. 

Beyond computational challenges, visual analytics systems that incorporate xAI have to present explanations in a way that users can easily interpret and trust. 
Providing details on the inner workings of the models may provide more transparency but runs the risk of overwhelming people with information that might lead to cognitive overload, where people struggle to process and understand the data.
This is in contrast to the goal of incorporating explainable AI with visual analytics. 
To mitigate these issues, a visual analytics design could explore alternative approaches to disseminate information in a multimodal way, enabling layered information without overwhelming people~\cite{jasimcommunityclick}. 
However, such approaches require interdisciplinary and collaborative efforts, combining expertise in AI, visualization, and human-computer interaction.

\subsection{Adoption Challenges}
Designing and developing visual analytics systems is one facet of adopting such systems in the civic domain.
However, long-term adoption presents unique challenges related to sustainability, economics, ethics, and technology integration.

\subsubsection{Lack of Visual Literacy across Stakeholders} 
Visual literacy, or lack thereof, persists as a challenge across many areas of visualization~\cite{mahyar2019civic}. 
It refers to the ability to extract, interpret, analyze, and communicate information through visualizations, including graphs, charts, maps, diagrams, and other visual media~\cite{burns2023we}. 
In the civic domain, a lack of visual literacy poses unique challenges across both the policymakers and public. 
For policymakers, despite being domain experts and experienced in civic activities, many are not adept at working with complex visualizations~\cite{mahyar2019civic}. 
Moreover, they often have to deal with a large amount of data, which, when converted into visualizations, the aggregation may reduce the granularity of underlying data and may not be able to capture nuances present in the data, leading to bias, misinterpretation, and ineffective decision-making~\cite{jasim2022supporting, jasim2023bridging}. 

The public also often lacks visual literacy, which necessitates developers to produce comprehensible visualizations that are easy to interpret. 
However, simplified visualizations can strip out the richness of the data and could lead to a skewed understanding. 
The flip side of this is the lack of inclusive approaches to building visualizations, which are difficult to interpret and are less useful in disseminating information to the general populace. 
Overcoming these challenges may require approaches beyond technological interventions by fostering visual literacy through education~\cite{firat2022interactive}, designing accessible and inclusive visualizations with the diverse audience in mind~\cite{mahyar2019civic, jasim2021communitypulse}, and exploring alternative interactions with visualizations beyond information delivery~\cite{mahyar2024reimagining}. 

\subsubsection{Lack of Infrastructure and Resources for Sustainability}

While research works can be demonstrated to be effective in addressing civic issues in a contained scope~\cite{kriplean2012supporting, zilouchian2015procid}, sustaining them for long-term adoption poses significant challenges due to infrastructural limitations, resource constraints, technical challenges, and evolving community needs. 
Many research projects heavily rely on short-term research grants or pilot programs without long-term financial plans, leading to abandoning the system alongside the project once the initial funding ends. 
In addition, with the passage of time, the technology becomes outdated, needing updates to continue support, which demands additional resources, including funding, programmers, and developers to maintain and continue operability of visual analytics systems. 
While open sourcing, crowdsourcing, and similar approaches can be explored to mitigate the resource challenges on a voluntary basis, without clear and scalable structuring, these approaches can at best maintain soon-to-be obsolete software, with few possibilities of scaling~\cite{reynante2021framework}. 
As such, these projects are often confined locally and rarely scale to larger municipal or national contexts. 

Beyond financial and technical challenges, civic technology must adapt to changing community needs while balancing data security and ethical standards. 
Visual analytics that address specific civic issues and demands today may become irrelevant without modifications and changes to adjust to evolving current affairs, social norms, political structures, and public opinions. 
To supplement the evolution of software programs, long-term data management and governance are needed to maintain historical records while adjusting to new data privacy standards~\cite{rustad2019towards}.  
In essence, successful and sustainable visual analytics systems in the civic domain require proactive planning, resource allocation, adaptive system design, and collaboration among stakeholders and developers. 

\subsubsection{Integration and Disruption in Established Workflows}
In many cases, introducing a novel technology such as visual analytics systems can be detrimental to the established procedure and workflow of civic engagement, public input analysis, and decision-making~\cite{coleman2001bowling, mahyar2019civic}. 
Government officials often rely on well-defined and documented methods for collecting and analyzing public input.
While these approaches might not be the most updated and technology-supported, introducing novel systems may create resistance among stakeholders accustomed to traditional processes.
Therefore, decision-makers are often resistant to integrate new technological interventions into existing frameworks due to these challenges~\cite{jasim2021communityclick}.
In addition, the visual analytics system introduces new dimensions to how public input can be analyzed to inform decisions. 
Collectively, such changes to the workflow might hinder rather than enhance public participation through established voluntary or crowdsourced approaches, as traditional approaches may not immediately translate to working with digital technologies. 
This might, in turn, demand more efforts, training, and commitment from the public and policymakers, rendering the idea of adopting new technologies rather unattractive. 

To mitigate the disruption of existing workflows, visual analytics systems can be integrated as a complement rather than a replacement~\cite{jasim2021communityclick, norman2015designx}. 
Such technologies can be gradually integrated as a fail-safe or an auxiliary method to enhance the current processes. 
This would allow for practitioners of established workflows to continue to follow their preferred methods while allowing for experimentation and gradual adoption of novel technologies in the long run. 
Positioning visual analytics systems as a supporting mechanism and fostering user agency over their usage will enable possibilities of adopting new technologies at the stakeholders' own pace, leading to a more natural transition between old and new approaches. 
Such approaches could also be beneficial in mitigating risks of misinterpretation due to limited exposure, building trust, and increasing the potential for sustained adoption. 

\subsubsection{Concerns about technology and AI}



While technology is becoming ubiquitous, in the civic domain, there are risks associated with relying solely on technology for engagement and analysis. 
Technological adoption may unintentionally favor individuals with greater access to technology, particularly younger users who are more familiar with digital tools, while older populations may be less inclined or proficient in using them~\cite{rogers2017understanding}.
Some studies argue that with careful design of user interfaces and experiences, the older population is equally, if not more likely, to adopt new technologies~\cite{while2024gerontovis, d2014three}. 
However, it does not overcome the concerns entirely, especially with the recent progress in AI following the introduction of disruptive technologies, including foundation models and their usage~\cite{danneels2004disruptive, medetalibeyoglu2025foundational}. 
To mitigate these challenges, it is crucial to design inclusive and accessible technologies that accommodate diverse user needs, provide training and support for less tech-savvy populations, and incorporate hybrid engagement strategies that blend digital and non-digital approaches to ensure equitable participation~\cite{d2014three, vaportzis2017older}.
It is also important to be transparent about technology and AI integration in civic spaces and design systems that are inclusive and adaptable to stakeholders' needs, rather than replacing them.
Furthermore, careful consideration is needed when integrating AI into visual analytics systems designed to support participatory democracy processes and disseminate information to avoid eroding already dwindling trust and avoid exacerbating the often tenuous relationship between the public and a technocratic government~\cite{sieber2024civic, novelli2024digital}.
Given the ephemeral landscape of AI-enabled technology and research, it is imperative to conduct longitudinal studies to identify nuanced challenges and potential solutions that can address long-term adoption challenges of technologies such as visual analytics systems that integrate AI. 

\subsubsection{Data Privacy and Ethical Considerations} 

All visualizations depend on data~\cite{chen2007handbook}. 
Visual analytics in the civic domain predominantly relies on public-generated input and feedback on issues concerning the community, city, or nation at scale. 
While visual analytics systems must be transparent to foster trust and accountability, balancing transparency with data privacy in the civic domain presents unique challenges, particularly in mitigating the risk of exposing sensitive or personally identifiable information such as location, identity, and social status.
Emergent computational technologies such as differential privacy~\cite{dwork2006differential, panavas2024illuminating}, widely considered a gold standard, can be used to address data privacy. 
On the other hand, manual moderation approaches could be undertaken to complement technological interventions.
However, implementing differential privacy requires significant computational resources and careful parameter tuning to balance privacy and data utility. 
Meanwhile, manual moderation, though valuable, introduces risks of misinterpretation, personal biases, and inconsistencies~\cite{gorwa2020algorithmic}.

Such challenges further introduce the potential for the misuse of public data and ethical concerns. 
In many cases, public-generated data in the civic domain is made publicly available~\cite{jasim2021communityclick, jasimcommunityclick}. 
If not properly anonymized, individuals' identities, opinions, or affiliations may become traceable, exposing them to risks such as surveillance, targeted political manipulation, and harassment.
From the researchers' and developers' perspective, they have to make extra effort to perform these tasks to avoid potential privacy leaks. 
Furthermore, biases in data collection and algorithmic processing can marginalize certain groups or reinforce existing inequalities.
It raises additional questions about the governance and ownership of civic data. 
It should be noted that in many instances, the policymakers manage and organize publicly generated data. 
However, the aforementioned issues might incur difficulties and apprehensions around collecting, analyzing, and disseminating public input, which might limit transparency, effectively enabling misinterpretation and leading to ineffective decision-making. 
Striking the right balance between open civic participation and data protection requires well-defined policies, privacy-preserving computational techniques, and adherence to ethical guidelines.

\section{Conclusion}

The transition of visual analytics systems from research prototypes to practical, widely adopted tools in the civic domain remains a persistent challenge. Through our experiences, we have identified key barriers, including organizational hierarchies, stakeholder misalignment, and resistance to new technologies, that hinder sustained adoption even when participatory design methods are employed. Our work underscores the need for deeper integration strategies that accommodate diverse stakeholder priorities and ensure long-term usability.
To bridge the gap between research and practice, future work should emphasize early stakeholder engagement, flexible adaptation to existing workflows, and sustained support for system evolution. Future research should also focus on developing frameworks for continuous collaboration, investigating methods for ensuring equity and accessibility, and exploring the role of AI-driven approaches in fostering interpretability and trust in visual analytics. Addressing these concerns will be crucial for making visualization research more impactful and actionable in real-world civic decision-making processes.

\noindent\textbf{Acknowledgment}

\noindent We would like to acknowledge our co-authors for their invaluable contributions to the projects discussed here over the past 10 years. We also thank all the local officials, community partners, and study participants who worked closely with us to inform the design, development, and evaluation of our visual analytics systems, and those who helped us deploy them in real-world civic contexts.


\balance

\bibliographystyle{eg-alpha-doi} 
\bibliography{egbibsample}       




\end{document}